\documentclass[conference]{IEEEtran}
\IEEEoverridecommandlockouts
\usepackage{cite}
\usepackage{amsmath,amssymb,amsfonts}
\usepackage{algorithmic}
\usepackage{graphicx}
\usepackage{textcomp}
\usepackage[table]{xcolor}
\usepackage[numbers]{natbib}

\usepackage{booktabs}
\usepackage{subcaption}
\usepackage{multirow}
\def\BibTeX{{\rm B\kern-.05em{\sc i\kern-.025em b}\kern-.08em
    T\kern-.1667em\lower.7ex\hbox{E}\kern-.125emX}}

\begin{document}

\title{
% Conference Paper Title*\\
Self-Knowledge Retrieval Augmented Generation Framework for Patent Matching
\thanks{*Corresponding Author}
}

\author{
    \IEEEauthorblockN{Jian Zhang$^{a}$, Songlin Lei$^{b}$, Zhuohao Yang$^{b}$, Bangli Liu$^{c}$, Ziwei Wang$^{c}$, Xufeng Weng$^{c}$, \\ Gehan Amaratunga$^{b}$, Yu Lin$^{b}$,  Hongwei Wang$^{abd*}$}
    \IEEEauthorblockA{$^a$ School of Computer Science and Technology, Zhejiang University, Hangzhou, China}
    \IEEEauthorblockA{$^b$ ZJU-UIUC Institute, Zhejiang University, Haining, China}
    \IEEEauthorblockA{$^c$ Shaoxing K3i Technology Co. Ltd}
    \IEEEauthorblockA{$^d$ State Key Laboratory of CAD\&CG, Zhejiang University, Hangzhou, China}
    \IEEEauthorblockA{\{jianzhang.22, hongweiwang\}@zju.edu.cn}
}

\maketitle

\begin{abstract}
Patent retrieval and matching based on large language models (LLMs) play a vital role in intellectual property protection. However, due to the complex structure of patent documents, dense technical terminology, and multi-modal information, traditional methods struggle to accurately identify subtle differences between patents. Existing LLM-based patent matching approaches typically rely on domain-specific pretrained or instruction tuning, which often entail high manual labeling costs and catastrophic forgetting. While retrieval-augmented generation (RAG) methods introduce external knowledge they fail to fully leverage LLM's capability to automatically parse patents and mine deep semantic relationships. To address these limitations, this paper proposes a self-knowledge RAG framework that guides LLMs to autonomously extract key technical entities and construct hierarchical ontological structures from patent matching queries, thereby enabling query expansion and precise retrieval. The method integrates the FAISS retrieval with a generative matching mechanism, leveraging self-knowledge to enhance the model's understanding of patent innovations and significantly improve retrieval and matching accuracy. Experimental results demonstrate the outstanding performance of the proposed method on real-world patent datasets, validating its effectiveness and application potential.
\end{abstract}


\section{Introduction}
Patent retrieval and matching is a critical issue in the field of intellectual property management \cite{patentManagement_1, patentManagement_2}. By identifying key information in patents and detecting patent documents with similar innovative concepts, it can effectively prevent intellectual property infringement \cite{infringement} . However, patent documents possess complex structures and diverse data types, often containing multimodal data such as text and images, along with extensive descriptions using specialized terminology. These characteristics pose significant challenges to patent retrieval \cite{challenges_1, challenges_2}. The intricate vocabulary descriptions make it increasingly difficult to discern subtle differences between patents, resulting in insufficient precision in patent matching.

With the emergence of large language models (LLMs), retrieval and matching tasks have seen rapid advancements, and patent retrieval has gained substantial attention. However, when applying these LLM technologies to the patent matching domain, challenges such as domain-specific vocabulary mismatch and difficulties in understanding technical terms and invention details persist \cite{llm_patent_challenges_1, pdc_3}. To mitigate these issues, current applications of LLMs primarily focus on domain-specific knowledge acquisition and matching. Specifically, recent approaches include pre-training LLMs on specialized corpora \cite{mozi} or constructing instruction-tuning datasets for specific tasks \cite{patentGPT}. Nevertheless, these methods typically require substantial human effort in corpus or dataset construction and carry the risk of catastrophic forgetting during training \cite{catastrophicForgetting}. An alternative strategy involves retrieval-augmented generation (RAG) using external knowledge \cite{rag, rag_patent_1, rag_patent_2}, leveraging finer-grained information to reduce noise in the patent retrieval process and enhance the accuracy of patent analysis. While parsing patents into triple-based knowledge bases as context for LLMs can improve patent document comprehension \cite{pdc_1, pdc_2, pdc_3}, such approaches often overlook the potential of LLMs to automatically parse patents and construct inter-patent relationships.

To address these challenges, we propose a self-knowledge RAG framework that strategically leverages LLMs to autonomously extract knowledge from patent matching queries. This knowledge subsequently guides both the retrieval process and the generative matching phase, transforming the LLM from a passive reasoning tool into an active analysis and information mining system. The core of our proposed method is a structured pipeline. LLM performs an in-depth analysis of the patent matching request, extracting key technical entities and constructing a hierarchical ontology (e.g., technology, function, application). The self-mined entities and ontological structures are used to form an enriched query, enabling more accurate retrieval of similar patents via the FAISS retriever \cite{faiss}. The original patent, retrieved similar patents, and self-mined entity-ontology structures are integrated into a concise yet information-rich instruction set. This allows the LLM to assimilate contextual patent knowledge, enhancing its comprehension of innovative aspects and improving the accuracy of search and generative matching.

\begin{figure*}
    \centering
    \includegraphics[width=0.8\linewidth]{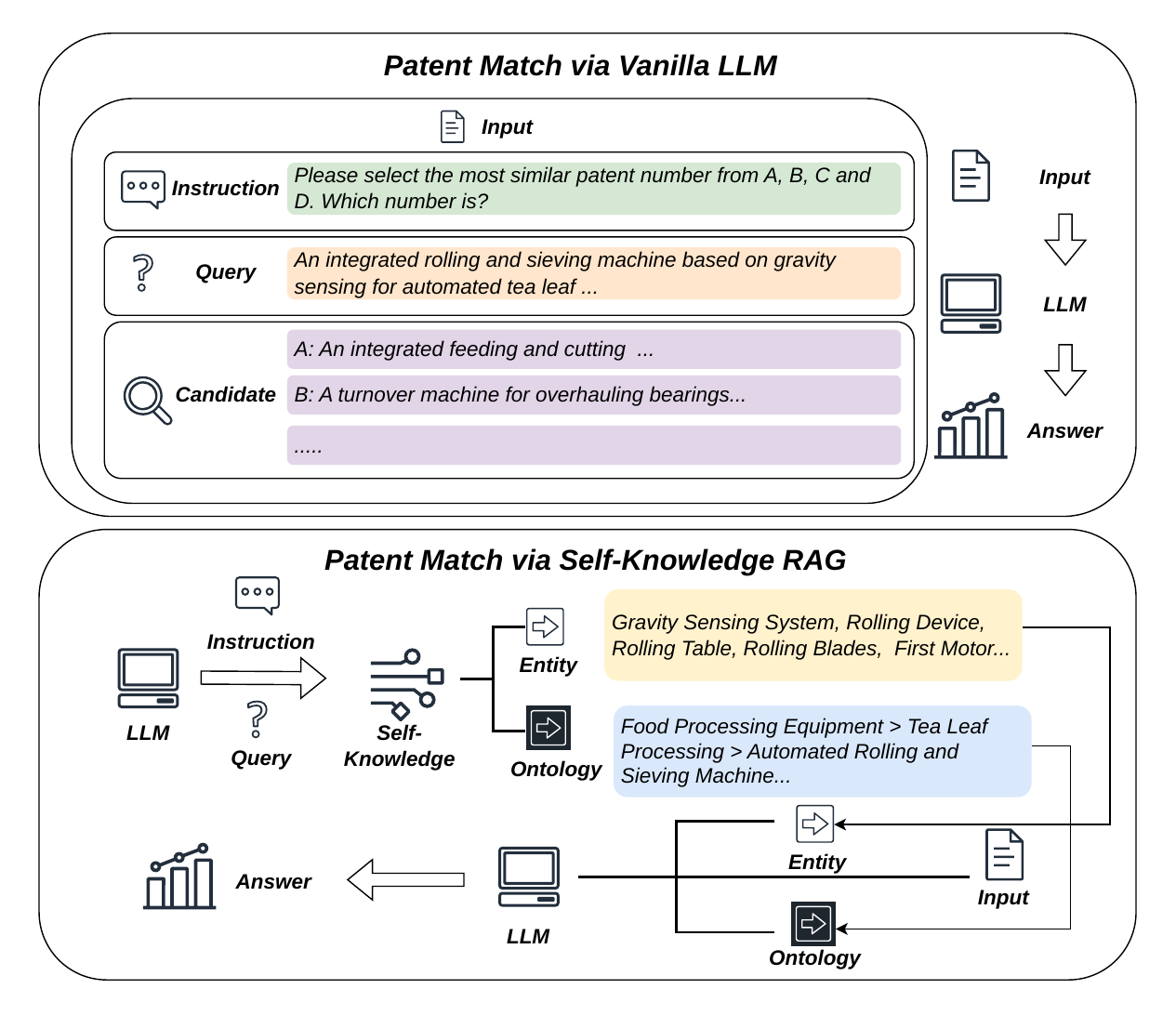}
    \vspace{-1em}
    \caption{The Proposed Framework of Self-Knowledge RAG Patent Matching}
    \label{fig:framework}
    \vspace{-1em}
\end{figure*}

The main contributions of this paper are as follows:
\begin{itemize}
    \item We designed a patent self-knowledge mining strategy based on LLMs, leveraging the model's capability to automatically extract entity and ontology information from patents.
    \item We introduced a self-knowledge guided RAG framework, which enhances FAISS vector index retrieval and generative matching through self-mined knowledge and query expansion, surpassing the performance of traditional methods such as Chain-of-Thought (CoT) and conventional RAG.
    \item The effectiveness of our proposed method was validated on real-world patent datasets, and case studies further demonstrate the advantages of our approach.
\end{itemize}

\section{Related Work}
Due to the strong diversity and structural complexity of patent documents, patent matching differs in focus from traditional patent retrieval. While patent retrieval concentrates on finding relevant patents, patent matching primarily addresses the similarity of technical innovations. Early patent matching methods mainly relied on keyword information \cite{keyinfo, keinfo_2} or learned patent innovation similarity through embedded spaces \cite{embedding_match, embedding_match_1, embedding_match_2}. These approaches failed to capture innovation similarities between patents, resulting in limited matching accuracy.

Recent patent matching research has primarily focused on frameworks constructed using LLMs, which leverage LLMs' powerful semantic understanding and emergent capabilities [50,29,61] to learn professional terminology and concepts through training. MoZi \cite{mozi} continuously trained LLMs on patent corpora and conducted instruction tuning for patent-related questions to enhance LLMs' understanding of technical details in patent documents. PatentGPT \cite{patentGPT} utilized external patent knowledge bases for pre-training to help LLMs capture relationships between patent entities. PatentGPT-Dense \cite{patentGPT-dense} further optimized human-reviewed matching behaviors through reinforcement learning from human feedback for matching alignment, demonstrating excellent performance in patent matching tasks.

Another strategy for utilizing large models adopts RAG framework \cite{rag_patent_3, rag_patent_4, rag} to enhance LLMs' domain-specific capabilities. However, some studies have shown that noise in retrieved documents may cause comprehension biases in LLMs, leading to knowledge conflicts \cite{knowledgeConflicts, knowledgeConflicts_2} and performance degradation \cite{performanceDegradation}. Some research has attempted to mitigate the impact of retrieval noise through graph-based RAG methods by refining retrieved documents to prevent LLM performance deterioration. Although these methods avoid domain-specific fine-tuning of LLMs and only use external knowledge such as knowledge triples to enhance response generation \cite{pdc_1, infringement, pdc_2, pdc_3}, these RAG-based approaches fail to fully utilize retrieved data for comprehension enhancement and don't effectively alleviate conflicts between LLMs' parametric memory and external knowledge.

\section{Method}
In the patent matching task, given a query patent $p_q$, the objective is to select the patent with high similarity of innovation from a set of $N$ candidate patents $\mathcal{P}=\{p_A, p_B, ..., p_N\} $, where each candidate patent $p_i$ is associated with a corresponding label $y_i$. The patent matching task involves constructing a query framework that outputs the identifier $y_i$ of the candidate patent most relevant to the target patent in terms of innovative description.
\begin{equation}
    \begin{aligned}
        P(\tilde{y}|p_{q},\mathcal{P})=\text{LLM}(p_{q},p_{\text{A}}\oplus\cdots\oplus p_{\text{D}})
    \end{aligned}
\vspace{-0.5em}
\end{equation}
% 其中，\oplus是连接操作符,表示给大模型输入专利内容p_q和各个候选专利p_A...p_D,大模型给出在候选专利P中与专利p_q最相似的标识符y_i
where operator $\oplus$ denotes a concatenation operation, the large language model $\text{LLM}$ takes query patent $p_{q}$ and candidate patent $\mathcal{P}$ as input, and outputs the identifier $y_i$  of the candidate patent most similar to  $p_{q}$.

This study proposes a self-knowledge RAG framework for patent matching based on mining with LLMs. The framework guides the LLM to extract key entities and ontological information from the target patent, constructing a structured knowledge system through self-mined information to enhance the accuracy and reliability of retrieval and matching. This approach eliminates traditional steps such as keyword extraction and matching in conventional retrieval tasks, adopting a three-stage paradigm of "Preparation-Retrieval-Reasoning." The overall framework of the method is illustrated in Fig \ref{fig:framework}.

\subsection{Self-Knowledge Mining from Query Patent}
\label{subsec:extraction}
This section introduces a self-mining approach that relies solely on LLMs to extract key information from query patent texts, without dependence on external knowledge bases. The method leverages large models to construct preliminary entity-ontology relationships.

Given the abstract of a target patent, the LLM is used to identify entities and ontologies. Specifically, customized instructions guide the model to extract all critical technical entities (e.g., specific technological concepts, component modules, and core methods) and ontological concepts related to patent technology from the abstract. Entities represent concrete technical elements extracted as keywords from the patent abstract. Mathematical representation of entity extraction is follows:
\begin{equation}
    \begin{aligned}
        P(v^{e}(p_q, P))&=\text{LLM}(\text{Instruct}_{e},p_q,P) \\
    \end{aligned}
\vspace{-0.5em}
\end{equation}

where $\text{Instruct}_{e}$ represents the entity extraction instruct for LLM. When provided with the instruction, the query patent, and candidate patents, the model generates entity lists $v^{e}(p_q, P)$ for both the query and candidate patents based on the given requirements.

The ontological information is primarily derived from the International Patent Classification (IPC) system, a hierarchical framework essential for organizing and categorizing patents based on technical innovation. This standardized classification system systematically groups patents across technical domains, enabling effective analysis and comparison of patents from diverse industries and languages. By leveraging this predefined conceptual system, abstract categories, attributes, and relationships are organized in a tree structure. This clarifies and systematizes the relationships between concepts, facilitating efficient construction of patent knowledge frameworks and semantic understanding of core patented technologies. The process of ontology construction can be formally expressed as:
\begin{equation}
    \begin{aligned}
        P(v^{o}(p_q, P))&=\text{LLM}(\text{Instruct}_{o}, p_q, V^{e}(p_q), P, V^{e}(P)) \\
    \end{aligned}
\vspace{-0.5em}
\end{equation}
where $\text{Instruct}_{o}$ represents instruct guides the construction of the ontology hierarchy for LLM. Using the input information, the model completes the ontology construction $v^{o}(p_q, P)$ for the query and candidate patents.

\subsection{Knowledge-guided Patent Retrieval}
\label{subsec:retrieval}
In this section, we leverage the self-mined knowledge obtained from the previous stage and enhance the retrieval process using the FAISS framework to achieve efficient similarity search.

The entity information acquired in Sec. \ref{subsec:extraction} are concatenated to form an expanded query string, which is then combined with the original abstract text. For example:
\begin{equation}
    \begin{aligned}
        \mathcal{T}_{\text{IR}}(V^{e}(p_{q}))=v^{e}_{1}(p_{q})\oplus\cdots\oplus v^{e}_{n}(p_{q}),
    \end{aligned}
\end{equation}
\vspace{-1.5em}
\begin{equation}
    \begin{aligned}
        p_q^{*} = p_q \oplus \mathcal{T}_{\text{IR}} 
    \end{aligned}
\vspace{-0.5em}
\end{equation}

where, $\mathcal{T}_{\text{IR}}(V^{e}(p_{q}))$ is expanded query patent entity information, $p_q^{*}$ is embeddings of expanded retrieval query patent.

This strategy significantly enriches the semantic information of the query, addressing potential issues such as incomplete expression or semantic sparsity in the original query.

Efficient vector similarity retrieval is implemented through the FAISS framework. A pre-trained language model (in this study, the BGE model is employed) encodes the abstract of each patent in the entire candidate patent library. These vectors are pre-built into a FAISS index, which is optimized for K-nearest neighbor (KNN) search in large-scale vector sets, enabling rapid retrieval of the most similar candidate patents. The pretrained language model encoding represents:

\begin{equation}
    \begin{aligned}
        h(p_{q}^{*}) &=\text{PLM}(p_{q}^{*}) \\
        h(p_{i}) &=\text{PLM}(p_i), \text{each } p_i \in P.
    \end{aligned}
\vspace{-0.5em}
\end{equation}
where $PLM(*)$ is pretrained language model for encoding retrieval information, $h(*)$ is hidden representation for retrieval information.

During the retrieval process, the expanded query string is transformed into a query vector using the same encoding model. The FAISS index then utilizes cosine similarity to identify the top-K candidate patents, generating a preliminary set of candidate patents. The similarity computation can be formulated as:
\begin{equation}
    \begin{aligned}
        C(p_{q}^{*},p_{i})=h(p_{q}^{*})\cdot h(p_{i}).
    \end{aligned}
\vspace{-0.5em}
\end{equation}
where $C(p_{q}^{*}, p_i)$ denotes top-K candidate patents list from similarity retrieval.

\subsection{Context-Aware Generation for Patent Match}
In the final decision stage, we integrate all available information into an instruction set rich with semantic context, enabling the LLM to perform deep reasoning and generate the final matching result. 

The constructed contextual instruction set primarily includes the following components:

System Instruction: Explicitly requires the LLM to assume the role of a patent examiner, conducting comprehensive matching based on all provided information with rigorous matching logic.

\paragraph{Query Patent}: The original abstract text of the target patent to be queried.

\paragraph{Self-Knowledge Mined Information}: Key entity information and ontological hierarchy obtained in Sec. \ref{subsec:extraction}.

\paragraph{Self-Knowledge Retrieval Context}: The top-K most similar patents retrieved through the FAISS framework in Sec. \ref{subsec:retrieval}. This provides the LLM with accurate sources of patent innovation information through similar contexts, enhancing its reasoning capability regarding patent innovativeness.

\paragraph{Task Instruction}: Requires the LLM to evaluate each candidate patent based on entity keywords and ontological hierarchy, considering similarities in technical domain, functionality, application, and component structure, to identify the patent with the highest similarity.

By inputting this combined instruction $\text{Instruct}_{rag}$ set into the LLM, which encompasses micro-level entity information, macro-level patent ontologies information, and rich contextual data, the model achieves more accurate matching results compared to conventional RAG methods or simple vector retrieval approaches.
The instruction construction can be formatted as:
\begin{equation}
    \begin{aligned}
        \mathcal{Z}_{rag} =  S(p_{q}^{*},p_{i}) \oplus p_q \oplus P \oplus V^{o}(p_q) \oplus V^{o}(P)
    \end{aligned}
\vspace{-0.5em}
\end{equation}
\begin{equation}
    \begin{aligned}
    P(A({p_q}, P))&=\text{LLM}(\text{Instruct}_{rag}, \mathcal{Z}_{rag}) \\
    \end{aligned}
    \vspace{-0.5em}
\end{equation}

where $P(A({p_q}, P))$ is answer of patent matching.

\section{Experiment and Analysis}

\subsection{Experiment Setup}
This section details the dataset, evaluation metrics, baseline models, and implementation details employed in our study.

\subsubsection{Dataset}
The dataset used in this paper employs PatentMatch to evaluate different patent matching capabilities. This dataset collects 1,000 patent matching instances from real patent documents, with detailed data statistics shown in Table \ref{tab:dataset}. The dataset contains 500 Chinese and 500 English patent entries, covering 8 categories defined by the International Patent Classification (IPC). To acquire more comprehensive patent innovation information through retrieval-augmented generation, we utilized a curated collection of 300,000 patents from \cite{memgraph}, using BGE-large as the retriever to obtain the top-k most relevant patents as input for the RAG method.

\begin{table}[]
    \centering
    \caption{Dataset Statistic of PatentMatch, including the count and proportion of data for each International Patent Classification (IPC). }
    % \vspace{-0.5em}
    \begin{tabular}{llrr}
    \toprule
    \textbf{IPC} & \textbf{Description} & \textbf{Count} & \textbf{Prop} \\
    \midrule
    HUM & Human Necessities & 304 & 30.4\% \\
    OPER & Performing Operations; Transporting & 264 & 26.4\% \\
    CHEM & Chemistry; Metallurgy & 60 & 6.0\% \\
    TEXT & Textiles; Paper & 26 & 2.6\% \\
    CONS & Fixed Constructions & 40 & 4.0\% \\
    MECH & Mechanical Engineering; Lighting;  & 100 & 10.0\% \\ 
    & Heating; Weapons \\
    PHYS & Physics & 160 & 16.0\% \\
    ELEC & Electricity & 46 & 4.6\% \\
    \midrule
    Total & & 1,000 & 100\% \\
    \bottomrule
    \end{tabular}
    % \vspace{1em}
    
    \label{tab:dataset}
    % \vspace{-2em}
\end{table}

\subsubsection{Evaluation Metrics}
Following prior work \cite{memgraph}, we adopt Accuracy (Acc) as the primary metric to assess the effectiveness of patent matching.

\subsubsection{Baselines}
We compare the following categories of baseline approaches:
Base Large Language Models: Qwen2-Instruct-7B, and GLM-4-Chat-9B, Qwen2.5-Instruct-14B.
Domain-specific LLMs: MoZi \cite{mozi}, PatentGPT \cite{patentGPT}, and PatentGPT-Dense \cite{patentGPT-dense}.
Chain-of-Thought (CoT) Reasoning \cite{cot}: This method uses structured prompts to guide models through step-by-step analysis of patent documents, enhancing comprehension of innovation points and improving answer generation.
Retrieval-Augmented Generation (RAG) \cite{rag}: This approach retrieves relevant patents as contextual information to assist LLMs in identifying patents with similar innovations.
To evaluate the individual contributions of CoT and RAG techniques, we applied each method separately to the tested base LLMs, enabling clear comparison of their relative effectiveness.

\subsubsection{Implementation Details}
In this study, we selected different LLMs as the backbone models, including Qwen2-Instruct-7B, GLM-4-Chat-9B, and Qwen2.5-Instruct-14B. During the entity generation and ontology generation phases, we constructed entity-ontology relationship graphs using carefully designed prompts. For the retrieval phase, we employed bge-large-v1.5 \cite{bge} as the encoder, utilizing language-specific models for Chinese and English respectively to generate word embeddings for cosine similarity computation. In RAG phase, we constructed corresponding RAG prompts and utilized the same backbone models that generated entities and ontologies for final output generation.

\subsection{Main Results}

\begin{table}[]
    \centering
    \caption{Main Results}
    \vspace{-0.5em}
    \begin{tabular}{lrrr}
    \toprule
    \textbf{Method} & \textbf{English} & \textbf{Chinese} & \textbf{Overall} \\
    \midrule
    \rowcolor{gray!20} \multicolumn{4}{l}{\textbf{SFT LLM}} \\
    MoZi-7B \cite{mozi} & 25.8 & 29.0 & 27.4 \\
    PatentGPT-1.5B \cite{patentGPT} & 26.2 & - & - \\
    PatentGPT-1.0-Dense-70B \cite{patentGPT-dense} & 66.2 & 72.0 & 69.1 \\
    \midrule
    % \rowcolor{gray!20} \multicolumn{4}{l}{\textbf{Backbone LLM}} \\
    % % Llama-3.1-Instruct-8B & 43.0 & 51.0 & 47.0 \\
    % Qwen2-Instruct-7B & 31.4 & 47.0 & 39.2 \\
    % GLM-4-Chat-9B & 66.4 & 65.0 & 65.7 \\
    % Qwen2.5-Instruct-14B & 63.8 & 70.0 & 66.9 \\
    % \midrule
    % \rowcolor{gray!20}\multicolumn{4}{l}{\textbf{Chain-of-Thought (CoT)}} \\
    % % Llama-3.1-Instruct-8B & 44.4 & 54.0 & 49.2 \\
    % Qwen2-Instruct-7B & 32.8 & 49.2 & 41.0 \\
    % GLM-4-Chat-9B & 68.0 & 65.6 & 66.8 \\
    % Qwen2.5-Instruct-14B & 64.0 & 71.2 & 67.6 \\
    % \midrule
    % \rowcolor{gray!20}\multicolumn{4}{l}{\textbf{Retrieval-Augmented Generation (RAG)}} \\
    % % Llama-3.1-Instruct-8B & 49.4 & 45.2 & 47.3 \\
    % Qwen2-Instruct-7B & 49.2 & 68.2 & 58.7 \\
    % GLM-4-Chat-9B & 75.8 & 69.4 & 72.6 \\
    % Qwen2.5-Instruct-14B & 70.8 & 64.2 & 67.5 \\
    % % \midrule
    % % \multicolumn{4}{l}{\textbf{MemGraph}} \\
    % % % Llama-3.1-Instruct-8B & 66.0\textsuperscript{†‡} & 64.6\textsuperscript{†‡} & 65.3 \\
    % % % Qwen2-Instruct-7B & 62.6 & 71.4 & 67.0 \\
    % % % GLM-4-Chat-9B & 82.8 & 80.8 & 81.8 \\
    % % % Qwen2.5-Instruct-14B & 75.8 & 75.0 & 75.4 \\
    % % Llama-3.1-Instruct-8B & \\
    % % Qwen2-Instruct-7B & 40.4 & 67.6 & 54.0 \\ % en 202 zh 338
    % % GLM-4-Chat-9B & 80.4 & 78.2 & 79.3 \\ % en 402 zh 391
    % % Qwen2.5-Instruct-14B & \\
    % \midrule
    % \rowcolor{gray!20}\multicolumn{4}{l}{\textbf{Ours}} \\
    % % Llama-3-Instruct-8B & \\
    % Qwen2-Instruct-7B & 34.4 & 69.8 & 52.1 \\ % en 172 zh 349
    % GLM-4-Chat-9B & 83.6 & 79.0 & 81.3 \\ % en 418 zh 395 
    % Qwen2.5-Instruct-14B & 79.2 & 82.2 & 80.7 \\ % en 396 zh 411
    % \bottomrule
    % \end{tabular}
    % \vspace{1em}
    \midrule
    \rowcolor{gray!20} \multicolumn{4}{l}{\textbf{Qwen2-Instruct-7B}} \\
    Vanilla LLM & 31.4 & 47.0 & 39.2 \\
    Chain-of-Thought (CoT) & 32.8 & 49.2 & 41.0 \\
    Retrieval-Augmented Generation (RAG) & \textbf{49.2} & 68.2 & \textbf{58.7} \\
    Ours & 34.4 & \textbf{69.8} & 52.1 \\
    \midrule
    \rowcolor{gray!20}\multicolumn{4}{l}{\textbf{GLM-4-Chat-9B}} \\
    Vanilla LLM & 66.4 & 65.0 & 65.7 \\
    Chain-of-Thought (CoT) & 68.0 & 65.6 & 66.8 \\
    Retrieval-Augmented Generation (RAG) & 75.8 & 69.4 & 72.6 \\
    Ours & \textbf{83.6} & \textbf{79.0} & \textbf{81.3} \\
    \midrule
    \rowcolor{gray!20}\multicolumn{4}{l}{\textbf{Qwen2.5-Instruct-14B}} \\
    Vanilla LLM & 63.8 & 70.0 & 66.9 \\
    Chain-of-Thought (CoT) & 64.0 & 71.2 & 67.6 \\
    Retrieval-Augmented Generation (RAG) & 70.8 & 64.2 & 67.5 \\
    Ours & \textbf{79.2} & \textbf{82.2} & \textbf{80.7} \\
    \bottomrule
    \end{tabular}
    \label{tab:main_results}
    \vspace{-2em}
\end{table}

Our proposed method was validated on a multilingual patent matching dataset and compared with multiple baseline models. As shown in Table 2, our approach demonstrates notable advantages in Chinese, English, and overall performance.

Among the vanilla LLM baselines, significant performance variations were observed across different models. PatentGPT-1.0-Dense-70B substantially outperformed smaller parameter models such as PatentGPT-1.5B and Mozi-7B, achieving an overall accuracy of 69.1. Among the three open-source backbone models, GLM-4-Chat-9B and Qwen2.5-Instruct-14B delivered the best results, with accuracy scores of 65.7 and 66.9, respectively.

After incorporating CoT strategy, all models showed performance improvements, with Qwen2-Instruct-7B exhibiting the largest gain—a 1.7\% increase over its baseline. In terms of single-language performance, GLM-4-Chat-9B also showed significant improvement on the English dataset, rising from 66.4 to 68.0. These results confirm that step-by-step reasoning via CoT helps LLMs better grasp innovative aspects of patents and effectively handle complexities in patent texts.

When the RAG strategy was applied, the three backbone models achieved further performance gains. Notably, Qwen2-Instruct-7B and GLM-4-Chat-9B showed marked improvements, with Qwen2-Instruct-7B reaching 68.2 on the Chinese dataset and an overall accuracy of 58.7. On the English dataset, GLM-4-Chat-9B achieved 75.8, with an overall score of 72.6.

Our proposed method consistently enhanced performance across both Qwen and GLM backbone models. The most pronounced improvement was observed with GLM-4-Chat-9B, which attained 83.6 on the English dataset and 79.0 on the Chinese dataset, resulting in an overall accuracy of 81.3—significantly surpassing all other backbone models. These outcomes indicate that our method effectively enhances LLM performance in patent matching tasks, demonstrating strong applicability and robustness, particularly in multi-lingual settings, thereby validating its effectiveness.

\subsection{Case Study}
This case study compares the prompts used in Vanilla LLM and our proposed method, as illustrated in Table \ref{tab:case_study}. The comparison reveals that the Vanilla LLM relies solely on limited information from query patent, leading to misinterpretations by LLM and ultimately incorrect matching results. In contrast, our proposed method leverages internally mined knowledge to enhance the model's comprehension of query patent, guiding it to correctly reason and identify the most accurately matched patent.

\begin{table*}[]
    \centering
    \caption{Case Study}
    \vspace{-0.5em}
    \begin{tabular}{p{17.5cm}}
    \toprule
    \rowcolor{gray!20}\textbf{Input Data} \\
    \textbf{Instruction}: Please select the most similar patent number from A, B, C and D. Which number is? Only choose from options A/B/C/D, without providing additional analysis. \\
    \textbf{Query Patent}: Socks and a manufacturing method therefor, ... and having the same appearance as an ordinary sock. \\
    \midrule
    \rowcolor{gray!20}\textbf{Vanilla LLM} \\
     \textbf{Answer}: \textcolor{red}{(False)} \\
     \hspace{2em} B. A foot state monitoring method, comprising: ... and hearth care guide information. \\
     \midrule
     \rowcolor{gray!20}\textbf{Self-Knowledge RAG} \\
     Refer to the following patent abstract to answer the subsequent question: \\
       \hspace{2em} Patent1:Socks and ... outer layer sock body; the inner layer sock body ... sleeving the toes and a connecting ... sock. \\
       \hspace{2em} Patent2:A foot odor-resistant sock,... left side of the sock body ... first heat insulating layer ... facilitates the healthcare of a user. \\
       \hspace{2em} Patent3:A sock structure ... the waterproof layer ... the first end (12) and the second end. \\
       \hspace{2em} Patent4:Socks, and in particular, ... is provided outside the pocket ... is an elongated hole, ...  and promotion value. \\
       \hspace{2em} Patent5:Disclosed are a shoe... preventing the foam layers ... a strong supporting performance ... the production efficiency. \\
    \textbf{Entity:} \\
        \hspace{2em} Inner Layer Sock Body, Outer Layer Sock Body, Toe Sleeve Part, Connecting Part, Toe Sleeves, Parallel Arrangement, Five-toed Sock, Breathability Improvement, Fungus Growth Prevention,  Manufacturing Method \\
    \textbf{Ontology:} \\
        \hspace{2em} \textit{Original abstract}: Apparel \> Footwear \> Socks with Special Features, \textit{Option A}: Apparel \> Footwear \> Orthopedic Socks, \textit{Option B}: Healthcare \> Monitoring \> Foot Temperature Monitoring, \textit{Option C}: Electronics \> Medical Devices \> Finger Probe, \textit{Option D}: Medical Devices \> Vascular Devices \> Artificial Blood Vessels \\
    \textbf{Answer:} \textcolor{red}{(True)} \\
     \hspace{2em} A. Disclosed are stockings for..., and gradually improve hallux valgus with long-term use. \\
    \bottomrule
    \end{tabular}
    \label{tab:case_study}
    \vspace{-2em}
\end{table*}

\section{Conclusion}
This paper proposes a novel patent matching framework that enhances the performance of RAG and LLM-based generative patent matching by mining intrinsic knowledge such as entities and ontologies from patents themselves. Our method effectively alleviates issues such as insufficient semantic understanding and poor retrieval accuracy in patent matching tasks, while also providing an interpretable approach to patent relevance by clarifying the rationale behind retrieval decisions. Future work will focus on dynamic ontology generation and integrating multimodal patent data (e.g., images and text) to further improve the model's adaptability to real-world patent retrieval and matching scenarios.

\section*{Acknowledgment}
We would like to thank the anonymous reviewers for their valuable comments. This work is supported by National Key Research and Development Program of China (2024YFF0907803).

\bibliographystyle{IEEEtranN}  
\bibliography{ref}

\end{document}